\documentstyle[prl,twocolumn,aps]{revtex}

\begin{document}
\draft
\title{Direct-interaction electrodynamics of a two-electron atom }
\author{Jayme De Luca}
\address{Universidade Federal de S\~{a}o Carlos, \\
Departamento de F\'{i}sica\\
Rod. Washington Luiz, km 235\\
Caixa Postal 676, S\~{a}o Carlos, S\~{a}o Paulo 13565-905}
\date{\today}
\maketitle

\begin{abstract}
We study numerically the dynamical system of a two-electron atom with the
Darwin interaction as a model to investigate scale-dependent effects of the
relativistic action-at-a-distance electrodynamics.\ This
dynamical system consists of a small perturbation of the Coulomb dynamics
for energies in the atomic range. The key properties of the Coulomb dynamics
are: (i) a peculiar mixed-type phase space with sparse families of stable
non-ionizing orbits and (ii) scale-invariance symmetry, with all orbits
defined by an arbitrary scale parameter. The combination of \ this peculiar
chaotic dynamics ((i) and (ii)), with the scale-dependent relativistic 
corrections (Darwin
interaction) generates the phenomenon of scale-dependent stability: We find
numerical evidence that stable non-ionizing orbits can exist only for a
discrete set of resonant energies. The Fourier transform of these 
non-ionizing orbits is a set of sharp frequencies. 
The energies and sharp frequencies of the non-ionizing orbits 
we study are in the quantum atomic range.
\end{abstract}


The Coulomb dynamical system of the helium atom is a very peculiar chaotic
system that exhibits Arnold diffusion\cite{Xia}, and with a typical
trajectory having an infinity of possible time-asymptotic final states. For
example, almost all negative-energy trajectories of Coulombian helium
display the generic phenomenon of ionization, namely, the ejection of one
electron\cite{Kaneko}. Several nonlinear dynamical systems share this
property of having more than one time-asymptotic final state, with the
respective basins for each outcome having a complicated structure in initial
condition space\cite{Bleher,Kandrup}. The numerical work on this paper is
based on stable Coulombian orbits of a two-electron atom that do not ionize
for several millions of turns of one electron around the nucleus. It is a
property of the Coulomb dynamics of a two-electron atom that most initial
conditions with a negative energy ionize very quickly in about 20 turns\cite
{Kaneko}. Then there are the very special initial conditions that do not
ionize due to a precise phase balance between the two electrons. These rare
non-ionizing orbits are defined very sharply in phase space and were first
studied in reference \cite{Kaneko} for plane orbits. Here we also develop a
numerical procedure to search for non-ionizing orbits among a large number
of possible tridimensional initial conditions.

The Coulomb Hamiltonian exhibits the scale invariance degeneracy: if we
scale time and space as $t\rightarrow Tt$ , $\vec{r}\rightarrow L\vec{r}$,
for $T^{2}/L^{3}=1$, the equations of motion are left invariant. For this
reason, the behavior of the Coulomb dynamics is the same in all scales, a
degeneracy which is broken by the relativistic effects of electrodynamics.
The phenomenon of breaking the scale invariance in electrodynamics was
explored analytically in \cite{PRL,long} for the Darwin interaction, which
is the low-velocity approximation to the Wheeler-Feynman
action-at-a-distance electrodynamics \cite{Anderson}. It was found in \cite
{PRL,long} that a simple resonant normal form approximation theory predicts
a discrete set of quantized scales very close to the quantum atomic
energies. Using these preliminary findings as guide, we present a numerical
investigation of the stability of non-ionizing orbits for the Darwin
dynamics and its dependence on the energy scale. It turns out that for
energies of atomic interest, the Darwin equations of motion approximate the
Coulomb equations plus a perturbation of size $\beta ^{2},$ with $\beta \sim
10^{-2}$. Therefore, non-ionizing stable orbits of the Darwin dynamics
should exist in the neighborhood of non-ionizing stable Coulombian orbits if
the perturbation does not force ionization. For these, our numerical results 
with the Darwin dynamics indicate that the non-ionizing property plus 
stabilty require sharply defined discrete energies. 

The Darwin interaction is not exactly a Lorentz invariant interaction\cite
{barcelona,Havas,Vishal}, so we study it as an approximation to the
relativistic action-at-a-distance electrodynamics, for the sake of including
the present approach into an underlying physical theory. Maxwell's theory
would seem to be the natural candidate for the comprehensive physical
theory, but it lacks time-reversibility and dipolar dissipation would forbid
the orbits studied in this paper. There is also the choice of other more
recent Lorentz-invariant Lagrangian\cite{Relativisticdirect} systems and 
constrained Hamiltonian dynamical systems\cite{Horace1,Horace2,Arcetri}, 
whose exact forms are actually more amenable to numerical treatment than the 
Wheeler-Feynman electrodynamics, but we shall not consider them here. The 
interested reader should consult reference \cite{Horace1}, where a
covariant approximation to Wheeler-Feynman electrodynamics is attempted
by the two-body Todorov equation of constraint dynamics.

This paper is organized as follows: in section I we review the state of the
art of the time-reversible action-at-a-distance electrodynamics, and if the
reader wants to skip this part the rest of the paper makes full sense as a
nonlinear dynamics study, except for the discussion at the end. In section
II we describe the numerical calculations with the Coulomb limit of the
Darwin interaction, find some non-ionizing orbits and their Fourier
transforms. In section III we include the scale dependent Darwin terms and
investigate the possibility of stable non-ionizing orbits. In section IV we
put the conclusions and discussion.

\section{Action-at-a-distance electrodynamics}

The Wheeler-Feynman\cite{Fey-Whe}electrodynamics developed from the
Schwarzschild-Tetrode-Fokker\cite{Schw-Tetr-Fokk} direct-interaction
functional. Equations of motion are derived from Hamilton's principle for
the action integral 
\[
S=-%
\mathop{\textstyle\sum}%
_{i}%
\textstyle\int%
m_{i}cds_{i}+e^{2}%
\mathop{\textstyle\sum}%
_{ij}%
\textstyle\int%
\textstyle\int%
\delta \left( \left\| x_{i}-x_{j}\right\| ^{2}\right) x_{i}\cdot
x_{j}ds_{i}ds_{j}, 
\]
where the four-vector $x_{i}(s_{i})$ represents the four-position of
particle $i$ parametrized by arc-length $s_{i}$ , double bars indicate
quadri-vector modulus $\left\| x_{i}-x_{j}\right\| ^{2}\equiv
(x_{i}-x_{j})\cdot (x_{i}-x_{j})$ and the dot indicates the usual Minkowski
relativistic scalar product of four-vectors.\ (integration is to be carried
over the hole particle trajectories, at least formally). The above action
integral describes an interaction at the advanced and retarded \ light-cones
with an electromagnetic potential given by half the sum of the advanced and
retarded Li\`{e}nard-Wierchert potentials \cite{Anderson}. Wheeler and
Feynman showed that electromagnetic phenomena can be described by this
direct action-at-a-distance theory in complete agreement with Maxwell's
theory as far as the classical experimental consequences\cite{Fey-Whe,Leiter}%
. This direct-interaction formulation of electrodynamics was developed to
avoid the complications of divergent self-interaction, as there is no
self-interaction in this theory, and also to eliminate the infinite number
of field degrees of freedom of Maxwell's theory \cite{Plass}. It was a great
inspiration of Wheeler and Feynman in 1945, that followed a lead of Tetrode 
\cite{Fey-Whe} and showed that with the extra hypothesis that the electron
interacts with a completely absorbing universe, the advanced response of
this universe to the electron's retarded field arrives {\sl at the present
time of the electron} and is equivalent to the local instantaneous
self-interaction of the Lorentz-Dirac theory\cite{Dirac}. The
action-at-a-distance theory is also symmetric under time reversal, as the
Fokker action includes both advanced and retarded interactions. Dissipation
in this time-reversible theory becomes a matter of statistical mechanics of
absorption\cite{Einstein}. The area of Wheeler-Feynman electrodynamics has
been progressing slowly but steadily since 1945: Quantization was achieved
by use of the Feynman path integral technique and the effect of spontaneous
emission was successfully described in terms of interaction with the future
absorber, in agreement with quantum electrodynamics\cite{Narlikar}. It was
also shown that it is possible to avoid the usual divergencies associated
with quantum electrodynamics by use of proper cosmological boundary
conditions\cite{Narlikar}. As far as understanding of the dynamics governed
by the equations of motion, the state of the art is as follows: The exact
circular orbit solution to the attractive two-body problem was proposed in
1946\cite{Schonberg} and rediscovered by Schild in 1962\cite{Schild}. The
1-dimensional symmetric two-electron scattering is a special case where the
equations of motion simplify a lot and it has been studied by many authors,
both analytically and numerically \cite{VonBaeyer,Driver1,Igor}. In this
very special case the initial value functional problem surprisingly requires
much less than an arbitrary initial function to determine a solution
manifold with the extra condition of bounded manifold for all times. It was
shown that the solution is uniquely determined by the interelectronic
distance at the turning point if this distance is large enough (this minimum
distance curiously evaluates to 0.49 Bohr radii by the action-at-a-distance
theory\cite{Driver1}, much larger than about one classical electronic radius
that one would naively guess). As a result of this theorem, there is a
single continuous parameter (the positive energy) describing the unique
non-runaway symmetric orbit at that given positive energy.

The Noether's four-constant of motion derived from the Fokker Lagrangian
involves an integral over the past history\cite{Anderson,Fey-Whe,Schild}.
For example in the case of a hydrogen atom this four-momentum constant
evaluates to \cite{Anderson} 
\begin{eqnarray*}
P^{\lambda } &=&m_{p}\dot{x}_{p}^{\lambda }+eA^{\lambda }(x_{p})+m_{e}\dot{x}%
_{e}^{\lambda }-eA^{\lambda }(x_{e}) \\
&&\!\!\!\!\!\!\!\!\!-2e^{2}\int_{\tau }^{\infty }d\tau _{p}\int_{-\infty
}^{\tau }d\tau _{e}\acute{\delta}\left( \left\| x_{e}-x_{p}\right\|
^{2}\right) (x_{p}-x_{e})\dot{x}_{e}\dot{x}_{p} \\
&&\!\!\!\!\!\!\!\!\!+2e^{2}\int_{-\infty }^{\tau }d\tau _{p}\int_{\tau
}^{\infty }d\tau _{e}\acute{\delta}\left( \left\| x_{e}-x_{p}\right\|
^{2}\right) (x_{p}-x_{e})\dot{x}_{e}\dot{x}_{p},
\end{eqnarray*}
where $\acute{\delta}$ represents the derivative of the delta function\cite
{Anderson}. Notice that because of this delta function, only finite portions
of the trajectory are involved: actually an extent of \ length $t\simeq
2r_{12}/c$ approximately. This non-local constant will behave very
differently from the local Coulombian energy, that is known to confine
orbits of a negative energy within a maximum separation distance. In the
case where the particles acquire a large separation (unbound state), the
hole past history is involved ($t\simeq 2r_{12}/c\simeq \infty $) in the
determination of the non-local energy constant.

As regards the mathematical structure of the equations of motion, for the
case of a two-electron atom the acceleration of electron 1 is given by\cite
{Hans} 
\begin{equation}
a_{1}(t)=\frac{-e}{m\gamma }\{E-\frac{v_{1}(t)}{c^{2}}E\cdot v_{1}(t)+\frac{%
v_{1}(t)}{c}\times B\},  \label{motion1}
\end{equation}
where $-e$ and $m$ are the electronic charge and mass, $\gamma \equiv $ $1/%
\sqrt{(1-\frac{v_{1}^{2}(t)}{c^{2}})^{1/2}}$and $E$ and $B$ are the\ total
electric and magnetic fields produced by electron 2 and the nucleus. In the
action-at-a-distance theory these fields are given by the average of the
retarded and advanced Li\`{e}nard-Wiechert fields, calculated with the
instantaneous position of the stationary nucleus and the retarded and
advanced positions of electron 2 at the times $t_{2}=t_{\mp }$ , which is
defined by the implicit condition

\[
R_{\mp }\equiv |r_{2}(t_{\mp })-r_{1}(t_{1})|=\mp c(t_{\mp }-t_{1}), 
\]
where the minus and plus signs are the conditions for the retarded and
advanced times respectively. The partial electric fields of electron 2
acting on electron 1 at time $t_{1\text{ }}$are \cite{Jackson} 
\begin{eqnarray*}
E_{-}(x_{1},x_{2-},v_{2-},a_{2-}) &=&\frac{-e(n_{-}-\beta _{2})}{\gamma
_{2}^{2}(1-n_{-}\cdot \beta _{2})^{3}R_{-}^{2}} \\
\!\!\!\!\!\!\! &&-\frac{e}{c}\left[ \frac{n_{-}\times \left\{ (n_{-}-\beta
_{2})\times \dot{\beta}_{2}\right\} }{(1-n_{-}\cdot \beta _{2})^{3}R_{-}}%
\right] ,
\end{eqnarray*}
where $R_{\mp }n_{\mp }\equiv r_{2}(t_{\mp })-r_{1}(t_{1})$ , $\beta
_{2}\equiv v_{2}/c,(-e),\gamma _{2}\equiv (1-\beta _{2}^{2})^{-1/2}$ and $c$
is the speed of light. The advanced field $E_{+}$ is obtained from the above
expression by replacing $t_{-}$ by $t_{+}$ and $c$ by $-c.$ The partial
magnetic fields of electron 2 are 
\[
B_{\mp }=\pm n_{\mp }\times E_{\mp }, 
\]
where the $\pm $ is to ensure an outgoing Poynting vector $(cE\times B)$ for
the retarded fields and an incoming Poynting vector for the advanced fields.
The total electric field in equation (\ref{motion1}) must include also the
instantaneous Coulomb electric field of the stationary nucleus.

Equation (\ref{motion1}) can suggest a paradox about causality, as the force
depends on the future of particle 2. In the following, and to finish this
introduction, we show that equation (\ref{motion1}), when written properly,
becomes a functional differential equation with {\em delayed argument only},
as first observed in\cite{Harvard}. To outline the essentials of the
explanation, let us first ignore the field of the nucleus and take the
nonrelativistic limit of (\ref{motion1})\ ($v_{1}=0$). In this
approximation\ the electric field $E$ entering in equation (\ref{motion1})
evaluates to $%
E=0.5E_{+}(x_{1},x_{2+},v_{2+},a_{2+})+0.5E_{-}(x_{1},x_{2-},v_{2-},a_{2-})$%
. Then we note that one can use equation (\ref{motion1}) as an equation of
motion for {\em particle 2 , }by solving the rearranged form of (\ref
{motion1}), 
\begin{eqnarray}
eE_{+}(x_{1},x_{2+},v_{2+},a_{2+}) &=&  \nonumber \\
&\!\!\!\!\!\!\!\!\!\!\!\!\!\!\!\!\!\!\!\!&\!\!\!\!\!\!\!\!\!\!\!%
\!-2ma_{1}(t)-eE_{-}(x_{1},x_{2-},v_{2-},a_{2-}),\!  \nonumber \\
&&  \label{causal}
\end{eqnarray}
for the most advanced acceleration of particle 2, $a_{2+}\equiv a_{2}(t+%
\frac{d_{+}}{c}).$ In the above form it is clear that the right hand side
involves only functions evaluated at times prior to the most advanced time,
defined by $s=t+\frac{d_{+}}{c}$, and no further advanced information is
necessary, eliminating the ghost of dependence on the future. In the same
way, the causal equation of motion of particle 1 is to be produced from the
equation for particle 2 by solving for the most advanced acceleration of
particle 1. For the special case of 1-dimensional motion of two electrons, $%
E_{+}=E_{+}(x_{1},x_{2+},v_{2+})$ depends only on the advanced velocity, and
(\ref{causal}) can easily be solved for this advanced velocity as a function
of the past history. In the 3-dimensional case there is an extra complexity,
as the acceleration appears in the Li\`{e}nard-Wiechert partial field $E_{+}$
in the form $n_{12}\times (n_{12}\times a_{2+})/r_{12}$. The bad news is
that the component of the acceleration along the advanced normal can not be
solved for from the value of the double-vector-product only. Because of this
degeneracy, equation (\ref{causal}) is an algebraic-differential equation,
and the null direction of the left hand side of (\ref{causal}) is a
constraint to be satisfied by the right hand side (the scalar product with 
$n_{12}$ must vanish). The numerically correct way to integrate this type of
equation is by use of the modern integrators for algebraic-differential
equations like DASSL \cite{Petzold} adapted for retarded equations (which
has never been done yet) or by dealing directly with the algebraic
constraint \cite{Nikitin-deluca}. According to the standard classification
of G. A. Kamenskii\cite{Elsgolts}, equation (\ref{causal}) belongs to the
class of differential-difference equations of neutral type. Even though more
complex, the motion is still causally determined by the past trajectory, as
we wanted to demonstrate, the price being an algebraic neutral delay 
equation.

As far as initial conditions go, the general theory on delay equations \cite
{Elsgolts} tells us that we need to provide an initial $C^{2}$ function
describing the position of particle 2 from $s-\frac{(d_{+}+d_{\_})}{c}=t-%
\frac{d_{\_}}{c}$ up to the initial instant $s=0$. The information on
particle 1 needed is also to be provided over twice the retardation lag seen
by particle 1. This is a short piece of trajectory for bound nonrelativistic
atomic orbits, but for a ionized state or a runaway orbit this can be the
whole past history! Unless further simplifications or conditions are added,
this is the generic problem at hand. The 3-dimensional cases of atomic
interest (e.g. helium) have never been studied, and they are more complex
than the 1-d scattering because one can have negative energy bound states
for example. Most relevant for physics is the question of the conditions for
the existence of a bounded manifold solution, which still needs to be
understood in the general case (it would be very curious if they turned out
to be a discrete set of negative energies). The only existing analytical
result in the 3-dimensional case is the linear stability of the
Schonberg-Schild circular orbits\cite{Hans}, resulting in an infinite number
of unstable solutions to the characteristic equation. The numerical treatment 
ofthe exact neutral equations displays instabilities and is generally 
difficult. In the following we resort to the Darwin approximation not
as much as a mathematical approximation to the action-at-a-distance
electrodynamics, but as a physical 
approximation of Lorentz-invariant dynamics in the atomic (shallow) 
energy range.

\section{Numerical Calculations for the Coulomb Dynamics}

To introduce our numerical calculations, we start from the scale-invariant
Coulomb limit of the Tetrode-Fokker-Wheeler-Feynman interaction: Let $-e$
and $m$ be the electronic charge and mass respectively and $Ze$ the nuclear
charge of our two-electron atom, which in this work is assumed to have an
infinite mass. All our numerical work uses a scaling which exploits the
scale invariance of the Coulomb dynamics: Given a negative energy, there is
a unique circular orbit at that energy with frequency $\omega _{o}$ and
radius$\ R$ related by $e^{2}/(m\omega _{o}^{2}R^{3})=1/(Z-\frac{1}{4}%
)\equiv \zeta (Z)$. We scale distance, momentum, time and energy as $%
x\rightarrow Rx$, $p\rightarrow m\omega _{o}Rp$ , $\omega _{o}dt\rightarrow
d\tau $ and $E\rightarrow m\omega _{o}^{2}R^{2}\hat{H}$, respectively. In
these scaled units, the Coulomb dynamics of the two-electron atom is
described by the scaled Hamiltonian

\begin{equation}
{\bf \hat{H}}=\frac{1}{2}(|\vec{p}_{1}|^{2}+|\vec{p}_{2}|^{2})+\zeta (Z)\{%
\frac{1}{r_{12}}-\frac{Z}{r_{1}}-\frac{Z}{r_{2}}\},  \label{Hamitwo}
\end{equation}
where $r_{1}\equiv |\vec{x}_{1}|,\ r_{2}\equiv |\vec{x}_{2}|,\ r_{12}\equiv |%
\vec{x}_{1}-\vec{x}_{2}|\ $(single bars represent euclidean modulus) and $%
\beta \equiv \omega _{o}R/c$. For a generic non-circular orbit, $\beta $\
plays the role of a scale parameter, and we recover the value of the energy
in ergs through $E=mc^{2}\beta ^{2}\hat{H}$. Notice that $\beta $ does not
appear in the scaled Hamiltonian, which is the scale invariance property.
From the scaled frequency $\hat{w}$ and scaled angular momentum $\hat{l}$ we
can recover the actual values in CGS units by the formulas 
\begin{equation}
w=\frac{mc^{2}\zeta (Z)\beta ^{3}}{e^{2}/c}\hat{w},\quad \qquad l=\frac{%
e^{2}/c}{\zeta (Z)}\frac{\hat{l}}{\beta }.  \label{scales}
\end{equation}
The only other analytic constant of the Coulomb dynamics, besides the energy
(\ref{Hamitwo}) is the total angular momentum, and this dynamics in chaotic
and displays Arnold diffusion, as proved in \cite{Xia} for a similar
three-body system.

The numerical calculations were performed using a 9th-order
Runge-Kutta embedded integrator pair\cite{Verner}. We chose the embedded
error per step to be $10^{-14}$, and after ten million time units of
integration the percentage changes in energy and total angular momentum were
less than $10^{-6}$. As a numerical precaution we performed the numerical
calculations using the double Kustanheimo coordinate transformation to
regularize single collisions with the nucleus\cite{Aarseth}. As these alone
are not enough for faithful integration, we checked that there was never a
triple collision, as the minimum inter-electronic distance was about $0.3$
units while the minimum distance to the nucleus was $\ 0.01$ units for all
the orbits considered in this work. We also checked that along stable
non-ionizing orbits we can integrate forward up to fifty thousand time
units, reverse the integration, go backwards another fifty thousand units
and recover the initial condition with a percentile error of $10^{-5}$. For
longer times this precision of back and forth integration degenerates
rapidly, which is due to the combined effect of numerical truncation and
stochasticity. The question of how far in time the numerical trajectories
approximate shadowing trajectories in the present system is far from trivial 
\cite{Grebogi}, but we assume it to be a time at least of the order of these
one hundred thousand units. (Energy conservation of one part in a million is
achieved for much longer times, even one billion time units).

The study of orbits of a two-electron atom was greatly stimulated by the
recent interest in semiclassical quantization, and these studies discovered
two types of stable zero-angular-momentum periodic orbits for helium ($Z=2$%
): the Langmuir orbit and the frozen-planet orbit \cite
{Wintgen,Alejandro}. A detailed study of the non-ionizing orbits of
Coulombian helium was initiated in reference \cite{Kaneko} for plane orbits,
and we describe some of their results below. There are basically two types
of non-ionizing orbits: \ Symmetric if $r_{1}=r_{2}$ for all times and
asymmetric if $r_{1}\neq r_{2}$ generically. Symmetric orbits are produced
by symmetric initial conditions like for example $x_{1}(0)=-x_{2}(0)$ and $%
v_{1}(0)=-v_{2}(0)$ or $x_{1}(0)=-x_{2}(0)$ and $v_{1}(0)=v_{2}(0)$ with $%
x_{1}(0)\cdot v_{1}(0)=0$ \cite{Wintgen} Because (\ref{Hamitwo}) is
symmetric under particle exchange, these orbits satisfy $r_{1}=r_{2}$ at all
times, and therefore cannot ionize if $H<0$ (both electrons would have to
ionize at the same time, which is impossible at negative energies). For
example the double-elliptical orbits (two equal ellipses symmetrically
displaced along the x-axis) discussed in \cite{long} are in this class.
Double-elliptical orbits are known to be unstable \cite{PRL,long} and we
find that they ionize in about one hundred turns because of the numerical
truncation error. Most symmetric plane orbits are very unstable to
asymmetric perturbations, with the exception of the Langmuir orbit for a
small range of $\ Z$ values around $Z=2$ \cite{Wintgen}$.$

The simplest way to produce an asymmetric non-ionizing plane orbit is from
the initial condition $\ x_{1}=(r_{1},0,0)$ , $\dot{x}_{1}=(0,v_{1}\sqrt{4/7}%
,0)$ , $x_{2}=(-1.0,0,0)$ , $\dot{x}_{2}=(0,-\sqrt{4/7},0)$, as suggested in 
\cite{Kaneko}. In Figure 1 we show the electronic trajectories for the
first three hundred scaled time units along a two-dimensional non-ionizing
orbit of $Ca^{+18}(Z=20)$ with $r_{1}=1.4$ and $v_{1}=1.28442$ in the above
defined condition. We used a numerical refining procedure to finely adjust $%
v_{1}$as to maximize the non-ionizing time and this condition of Figure 1
does not ionize for one million time units. The orbit survives that far only
for a very sharp band of values of $v_{1}$, other neighboring values
producing quick ionization. This orbit was named double-ring torus in \cite
{Kaneko}. The other possible type of non-ionizing orbit resulting from the
above initial condition, depending on\ $(r_{1},v_{1}),$ is what was named
braiding torus in reference \cite{Kaneko}, with both electrons orbiting 
within the same region. A search over $(r_{1},v_{1})$ was conducted in \cite
{Kaneko}, and it was found that most values of $(r_{1},v_{1})$ produce quick
ionization except for a zero-measure set of $(r_{1},v_{1})$ values where
braiding tori or double ring orbits are found. This suggests the general
result that non-ionizing orbits are rare in phase space.

To search for general tridimensional non-ionizing orbits in phase space, it
is convenient to introduce canonical coordinates $\vec{x}_{d}$ and $\vec{x}%
_{c}$

\begin{eqnarray}
\vec{p}_{d} &\equiv &(\vec{p}_{1}-\vec{p}_{2})/\sqrt{2},\qquad \vec{x}%
_{d}\equiv (\vec{x}_{1}-\vec{x}_{2})/\sqrt{2}  \nonumber \\
\vec{p}_{c} &\equiv &(\vec{p}_{1}+\vec{p}_{2})/\sqrt{2},\qquad \vec{x}%
_{c}\equiv (\vec{x}_{1}+\vec{x}_{2})/\sqrt{2}.  \label{Cchange}
\end{eqnarray}
Initial conditions with $\vec{x}_{c}=$ $\vec{p}_{c}=0$ describe
double-elliptical orbits (and circular as a special case). To generate an
elliptical initial condition, we exploit the scale invariance and set the
energy to minus one. It is easy to check that elliptical orbits of the
Hamiltonian (\ref{Hamitwo}) with an energy of minus one must have a total
angular momentum of magnitude ranging from zero to two. To exploit the
rotational invariance of (\ref{Hamitwo}), we can choose the plane defined 
at $\vec{x}_{c}=$ $\vec{p}_{c}=0$ by the angular momentum $%
\vec{L}=$ $\vec{x}_{d}\times \vec{p}_{d}+\vec{x}_{c}\times \vec{p}_{c}=\vec{x%
}_{d}\times \vec{p}_{d}$ to be the $xy$ plane. On this $xy$ plane\ a single
number $0<|\vec{x}_{d}\times \vec{p}_{d}|<2$ (the angular momentum),
determines completely the elliptical orbit. The next step in producing a
generic orbit is to add all possible perturbations along $\vec{x}_{c}$ and $%
\vec{p}_{c}$ to the chosen elliptical orbit. These are six directions and
once we are looking for bound oscillatory orbits, we can choose $z_{c}=0$,
once $z_{c}$ has to cross the $xy$ plane\ at some point. These are five
numbers to vary and plus the angular momentum of the elliptical orbit it
totals six parameters. Our numerical search procedure consists in varying
these six parameters over a fine grid, integrating every single initial
condition until the distance from one electron to the nucleus is greater
than twenty units, which is our ionization criterion. This criterion fails
if the orbit has a very low angular momentum because these can go far away
from the nucleus and come back, and therefore our search possibly misses
low-angular-momentum non-ionizing orbits. As the majority of the initial
conditions ionize very quickly, this search procedure is reasonably fast. We
first perform a coarse search for ionization times above one thousand units
and then refine in the neighborhood of each surviving condition to get
conditions that do not ionize after one million time units.

Using the above numerical search procedure we found the tridimensional
non-ionizing initial condition of Figure 2 for helium, 
 a tridimensional double-ring orbit generated by the initial condition
\begin{eqnarray*}
x_{1} &=&(1.2812617,0.0147169,0.0) \\
x_{2} &=&(-1.5511484,0.0147169,0.0) \\
p_{1} &=&(-0.0194868,0.4398889,0.1094930) \\
p_{2} &=&(-0.0194868,-0.7972467,0.1094930),
\end{eqnarray*}
which does not ionize before ten million turns. 
(After the search and refinement, we scaled this orbit's energy to minus one, 
for later convenience). 
We also found the non-ionizing orbit orbit of Figure 3 for H-minus 
($Z=1$), a tridimensional orbit generated by the condition
\begin{eqnarray*}
x_{1} &=&(1.9776507,-0.3411364,0.0) \\
x_{2} &=&(-1.2288121,-0.3411364,0.0) \\
p_{1} &=&(0.0421302,0.5057782,0.2810539) \\
p_{2} &=&(0.0421302,-0.4132970,0.2810539),
\end{eqnarray*}
which does not ionize before one million turns (Coulombian energy of this
condition is also minus one). This last orbit is fragile 
and numerically harder to find: as the first
electron has an orbit very close to the positive $Z=1$ charge, there remains
only a dipole field to bind the second electron. As the outer electron
is much slower in the scaled units, we had to plot the
first $10000$ time units of evolution to display the generic features
of the trajectory. Non-ionizing orbits of $%
H^{-}$ are very rare in phase space, which is reminiscent of the quantum
counterpart, as the $H^{-}$ ion is known to have only one quantum bound
state at $E\simeq -0.55mc^{2}\alpha ^{2}$, very close to the ionization
threshold $(-0.5mc^{2}\alpha ^{2})$\cite{hminus}.

One remarkable fact about these non-ionizing orbits is that they all have a
very sharp Fourier transform. This property makes them approximately
quasi-periodic orbits.\ For example in Figure 4  we plot the fast Fourier
transform of the orbit of Figure 2, performed using $\ 2^{16}$ points. (It
seems that there are at least two basic frequencies in the resonance
structure of Figure 4). Even though these orbits look like quasi-periodic
tori, there seems to be a thin stochastic tube surrounding each orbit, as
evidenced by a small positive maximum Lyapunov exponent. We calculated
numerically this maximum Lyapunov exponent by doubling the integration times
up to $T=10^{7}$ and found that the exponent initially decreases but then
saturates to a value of about $0.001$ for the orbits of Figures 1, 2 and
3. The gravitational three-body problem has recently been proved to display
Arnold diffusion\cite{Xia}, and this numeriacally calculated positive 
Lyapunov exponent suggests that the same is true for the two-electron 
Coulombian atom.

\section{Numerical calculations for the Darwin Dynamics}

The numerical integrations in this section are performed using the Darwin
approximation. The Darwin equations of motion are a $\beta ^{2}$
perturbation of the Coulomb dynamics, of size $\beta ^{2}\sim 10^{-4}$ for
atomic energies. In the scaled units of section II the Darwin Hamiltonian is
the following $\beta ^{2}$ perturbation of Hamiltonian (\ref{Hamitwo}) 
\begin{eqnarray}
{\bf \hat{H}}_{D} &=&\frac{1}{2}(|\vec{p}_{1}|^{2}+|\vec{p}_{2}|^{2})+\zeta
(Z)\{\frac{1}{r_{12}}-\frac{Z}{r_{1}}-\frac{Z}{r_{2}}\}  \nonumber \\
&&-\frac{\zeta (Z)\beta ^{2}}{2r_{12}}[\vec{p}_{1}\cdot \vec{p}_{2}+(\hat{n}%
_{12}\cdot \vec{p}_{1})(\hat{n}_{12}\cdot \vec{p}_{2})]  \nonumber \\
&&-\frac{\beta ^{2}}{8}[|\vec{p}_{1}|^{4}+|\vec{p}_{2}|^{4}],  \label{Darwin}
\end{eqnarray}
where $\hat{n}_{12}\equiv (\vec{x}_{1}-\vec{x}_{2})/r_{12}$. The second line
represents the Biot-Savart magnetic interaction plus the first relativistic
correction to the static electric field and the last line describes the
relativistic mass correction. Notice that these are both proportional to the
small parameter $\beta ^{2}$, which makes them a small scale-dependent
perturbation on the scale invariant Coulomb Hamiltonian (first line). It is
possible to regularize the Darwin equations with the same double-Kustanheimo
transformation\cite{Aarseth}, only that here one needs to define the
regularized time using the higher powers $dt=r_{1}^{2}r_{2}^{2}ds$, instead
of the lower powers\ $dt=r_{1}r_{2}ds$ used to regularize the Coulomb
equations\cite{Aarseth}.

The main question we address numerically in this section is the dependence
of the stability of a non-ionizing orbit with the energy scale of the orbit
. Here we use the word stability to mean ionization-stability: We call an
initial condition ionization-stable if any small perturbation of it produces
another non-ionizing orbit. The scale-dependent Darwin terms (of size $\beta
^{2})$ produce significant deviations from the Coulomb dynamics only in a
time-scale of order $1/\beta ^{2},$ which we find numerically to be the
typical time for a non-ionizing Coulombian initial condition to ionize along
the Darwin vector field. This poses a numerical difficulty if $\beta $ is
too small because one has to integrate the orbit for very long times to
investigate the stability. It turns out that ionization-stable orbits can be
found at larger values of $\beta $ for larger values of $Z$ . Here the
dynamical stability mechanism is reminiscent of quantum atomic physics,
where the values of $\beta $ vary with the nuclear charge as $\beta \sim
Z/137$. Large values of $Z$ facilitate the numerical procedure and in the
following we present the numerical investigation of the stability of
non-ionizing orbits starting from the large $Z$ case.

Let us start with the $\ Z=20$ calcium ion two-electron system along the
non-ionizing orbit of Figure 1 by fixing $r_{1}=1.4$ and $v_{1}=1.28442.$ in
the condition defined in section II. To test the stability of the orbit at
each value of $\beta $ we add a random perturbation of average size $\beta
^{2}$ to the initial condition and integrate the Darwin dynamics until
either we find ionization or the time of integration is greater than $10^{7}$
time units We repeat this for at least twelve randomly chosen perturbations
(because of the twelve degrees of freedom) and the minimum time to
ionization is plotted in Figure 5 as a function of $\beta $. It can be seen
that only for a narrow set of values around $\beta \sim 0.037$ this minimum
time to ionization was greater than $10^{6}$. or the other values it
decreases rapidly to a value of about $10^{3}$. One could argue that for the
other values of $\beta $ the non-ionizing initial condition has shifted away
from the $v_{1}=1.28442$ initial condition and this being the reason that
our orbit ionized. To test this, we fixed $\beta $ at a ''bad '' value for
example $\beta =0.02$  and varied the plane initial condition in the
neighborhood of this condition of Figure 1. We found that the minimum time
to ionization was always about $10^{3}$ (also the maximum time before
ionization was about $10^{3}$). We also searched in a bigger neighborhood, 
of size proportional to $\beta$. This suggests the interpretation that 
for the special resonant value of $\beta =0.037$ the net diffusive effect 
of the scale-dependent term
vanishes, allowing a non-ionizing perturbed manifold. In order to have a 
direct interpretation (in atomic units) of the scale parameter $\beta$, 
it is convenient to scale to minus one the energy of the initial condition 
of Figure 1 (by exployting the Coulombian scale invariance). After this,
 the energy of the orbit in 
ergs evaluates to $ E=mc^{2}\beta ^{2}\hat{H}=-mc^{2}\beta ^{2}$, and for 
$\beta =0.037$ this is approximately $-24.59$ atomic units. The total 
angular momentum of this
orbit is $l_{z}=7.94\hbar $. This orbit's energy is above the ionization 
continuum of the ion, $E=-mc^{2}\alpha ^{2}Z^{2}/2=-200$ atomic units, 
but it is still in the quantum range. It serves nevertheless to demonstrate 
that this dynamical system might exhibit non-ionizing stable orbits only at 
very sharply defined energy values.

For the orbits of Figures 2 and 3, the above procedure becomes prohibitively
slow, as the value of $\beta$ are much smaller and one must integrate for
very long times, much beyond the estimated shadowing time. To partially 
overcome this we used a larger amplitude random perturbation 
(of average size $20\beta^2$), to produce faster ionization.
The drawback with this is that the minimum ionization time does not show
pronounced peaks, only the average ionization time still showing a signature
of scale dependence. In Figure 6 we show this average time for the orbit of 
Figure 3. This property of sharply defined energies can possibly be found 
for the lower-lying energies below the ionization threshold as well. 
These orbits would involve configurations where the electrons come very 
close to the nucleus and acquire a large velocity. Even though our 
integrator is regularized, the correct physical electronic repulsion is 
greatly amplified when one electron has a relativistic velocity and the 
Darwin approximation can not describe the physics then. Actually, it is 
known that the Darwin interaction can produce unphysical effects when pushed 
to relativistic energies\cite{Bessonov}. We therefore do not expect to 
find these low-lying atomic energy scales with the present Darwin 
approximation and shall be contempt with these interesting result 
already.

For the same reason given above, we do not study here the frozen-planet
periodic orbit (the two electrons performing one-dimensional periodic motion
on the same side of the nucleus, with the inner electron rebounding from the
origin, an artifact of regularization). The main problem being the failure
of the Darwin approximation, as the inner particle goes to the speed of 
light\cite{Bessonov}. The correct relativistic dynamics can actually 
produce a new\ {\em physical} inner turning point very close to the origin 
but not {\em at} the origin as the regularized motion, and we discuss 
elsewhere\cite{Inner}.

Last, we consider the non-ionizing symmetric periodic orbit called the
Langmuir orbit, where the two electrons perform symmetric bending motion
shaped approximately like a semi-circle\cite{Quantlang}. For the Coulomb
two-electron atom with $Z=2$ this orbit was found to have a zero maximum
Lyapunov \ exponent\cite{Wintgen}. The orbit is therefore neutrally stable,
which is the best one can expect from a periodic orbit of a Hamiltonian
vector field. (Absolute stability violates the symplectic symmetry, which
says that to every stability exponent $\lambda $ one should have a $%
1/\lambda $ exponent). It is a simple matter to obtain the Langmuir-like
orbit for the Darwin Hamiltonian at any given value of $\beta $: all it
takes is a little adjusting in the neighborhood of the Coulombian Langmuir
condition. We attempted to investigate numerically any scale-dependent 
diffusion away from this Darwin-Langmuir condition for $\beta $ in 
the atomic range, but again the numerics is prohibitively slow at the time 
of writing this work.

\section{Conclusions and Discussion}

The simplified dynamical mechanism behind resonant non-ionization seems 
to go intuitively as follows: The peculiar scale-invariant Coulomb dynamics
determines the non-ionizing orbits within narrow ''stochastic tubes ''. The
next step is the action of the small scale-dependent relativistic
corrections that produce a slow diffusion of the orbit out of the thin tube
in a time of the order of $1/\beta ^{2}$. After this, quick ionization
follows. Only at very special resonant values of $\beta $ the relativistic
terms leave the orbit within the tube, a resonant effect that depends on $%
\beta $, fixing the energy scale. In the literature, the escape to infinity
from simpler to understand two-degree-of-freedom systems has been attributed
to cantori, which, as is well known, can trap chaotic orbits near regular
regions for extremely long times\cite{Kandrup}. In the present larger
dimensional case it appears that resonances are also controlling the escape
to infinity of one electron by the existence of extra resonant constants of
motion\cite{PRL,long}. This seems to be in agreement with the numerical
results of very sharp peaks for the minimum ionization time. We have tried
to concentrate on the physics described by this combination of chaotic
dynamics on a two-electron atom with inclusion of relativistic correction,
while discussing this highly nontrivial result of nonlinear dynamics.

In references \cite{PRL,long} we noticed that a simple resonant normal form
criterion gives a surprisingly good prediction for the discrete atomic
energy levels of helium. The resonant structure was calculated using the
Darwin interaction (\ref{Darwin}), which is the low-velocity approximation
to both Maxwell's \cite{PRL,long} and Wheeler-Feynman's\cite{Anderson}
electrodynamics. As we saw in section II, the Coulombian non-ionizing orbits
are far from circular, and these orbits would radiate even in dipole
according to the time-irreversible Maxwell's electrodynamics (circular
orbits radiate only in quadrupole but are linearly unstable). It becomes
then clear that the heuristic results of \cite{PRL,long} can only have a
physical meaning in the context of a time-reversible theory (as the
action-at-a-distance electrodynamics for example).

The combination of chaotic dynamics with relativistic invariance has never
been explored numerically, and most known Lorentz-invariant dynamical
systems are for one particle and possess trivially integrable dynamics. The
situation gets unexpectedly much more complicated for more than one particle
(apart from the trivial non-interacting many-particle system): Due to the 
famous no-interaction theorem\cite{Sudarshan}, the relativistic description 
of two directly interacting particles is impossible within the Hamiltonian
formalism and its set of ten canonical generators for the Poincare group
\cite{Arcetri}. Description of interacting particles is possible only in 
the context of constraint dynamics, with eleven canonical generators and 
with the Dirac bracket replacing the Poisson bracket. 
For example the relativistic action-at-a-distance equations for two 
interacting electrons are non-local and possess only infinite-dimensional 
constrained Hamiltonian representations\cite{Igor,Single-time}. The 
interested reader should consult some recently found two-body 
direct-interaction relativistic Lagrangian dynamical 
systems\cite{Relativisticdirect} as well as the constraint-dynamics 
direct-interaction models recently used in chromodynamics and two-body 
Dirac equations\cite{Horace1,Horace2,Arcetri}. The nonlinear 
dynamics of these models could display interesting and so far inexplored 
dynamical behaviour.

It would be natural to wonder if one can find an analogous scale-dependent 
dynamics for a dynamical system describing the hydrogen atom,
apparently the simplest example of Lorentz-invariant two-body relativistic
dynamics of atomic interest. It turns out that hydrogen is not simpler than
helium at all, but it appears to us that there is an essential difference
which has actually made the interesting dynamics of a two-electron atom
amenable to study already within the Darwin approximation: In a two-electron
atom orbits with a negative energy can ionize, while in hydrogen this might
be possible only if one includes all orders of the relativistic
action-at-a-distance interaction. (As we saw in section I, the \ ''Noether's 
energy constant'' involves a segment of the past trajectory, and a negative 
value does not forbid ionization). Ionization with a negative energy would be
impossible for hydrogen within the Darwin approximation (unless the electron
goes to the speed of light). This is indication
that in hydrogen the essential physics described by the action-at-a-distance
electrodynamics is of non-perturbative character. The paradoxical result of
the infinite linear instability of circular orbits in atomic hydrogen\cite
{Hans} is another warning of this non-perturbative dynamics. 

\section{Acknowledgments}

The author acknowledges the support of FAPESP, proc. 96/06479-9 and CNPQ,
proc. 301243/94-8(NV).

\begin{figure} 

\caption{Non-ionizing double-ring orbit for $Ca^{+18}$ ion ($Z=20$), 
obtained from the initial condition  
$\ x_{1}=(r_{1},0,0)$ , $\dot{x}_{1}=(0,v_{1}\sqrt{4/7}%
,0)$ , $x_{2}=(-1.0,0,0)$ , $\dot{x}_{2}=(0,-\sqrt{4/7},0),$ 
 with $r_{1}=1.4$ and $v_{1}=1.28442$ 
, trajectories are shown for the first 300 time units, the inner ring 
represents the orbit of electron 1 and the outer ring represents the 
orbit of electron 2, the units are the scaled units defined in 
section II. }    
\end{figure}

\begin{figure} 

\caption{Non-ionizing double-ring tridimensional orbit for helium
($Z=2$), trajectories are shown for the first 200 time units, the inner 
ring represents the plane projection of the orbit of electron 1 and 
the outer ring represents the projection of the orbit of electron 2. 
Positions are in the scaled units defined in section II. }    
\end{figure}

\begin{figure} 

\caption{Non-ionizing tridimensional orbit for H-minus
($Z=1$), trajectories are shown for the first 10000 time units. The 
inner ring represents the plane projection of the orbit of electron 1 
and the outer ring represents the projection of the orbit of electron 2.
Trajectory of the (fastest) electron 1 winds almost everywhere in the 
the dark inner core of figure.  
Positions are measured in the scaled units of section II. }
\end{figure}

\begin{figure} 

\caption{Fast Fourier Transform of the orbit of Figure 2 using $2^{16}$
points. Frequencies 
are measured in the scaled units of section II.}     
\end{figure}

\begin{figure} 

\caption{Minimum time to ionization (among 24 random perturbations of 
average size $\beta^2 $ added to the orbit of Figure 1)
$\beta$ is the adimensional parameter and
time is measured in the scaled units of section II.}      
\end{figure}

\begin{figure} 

\caption{Average time to ionization (among 12 random perturbations of 
average size $20 \beta^2 $ added to the orbit of Figure 3)
 $\beta$ is the adimensional scale parameter and
time is measured in the scaled units of section II.}      
\end{figure}

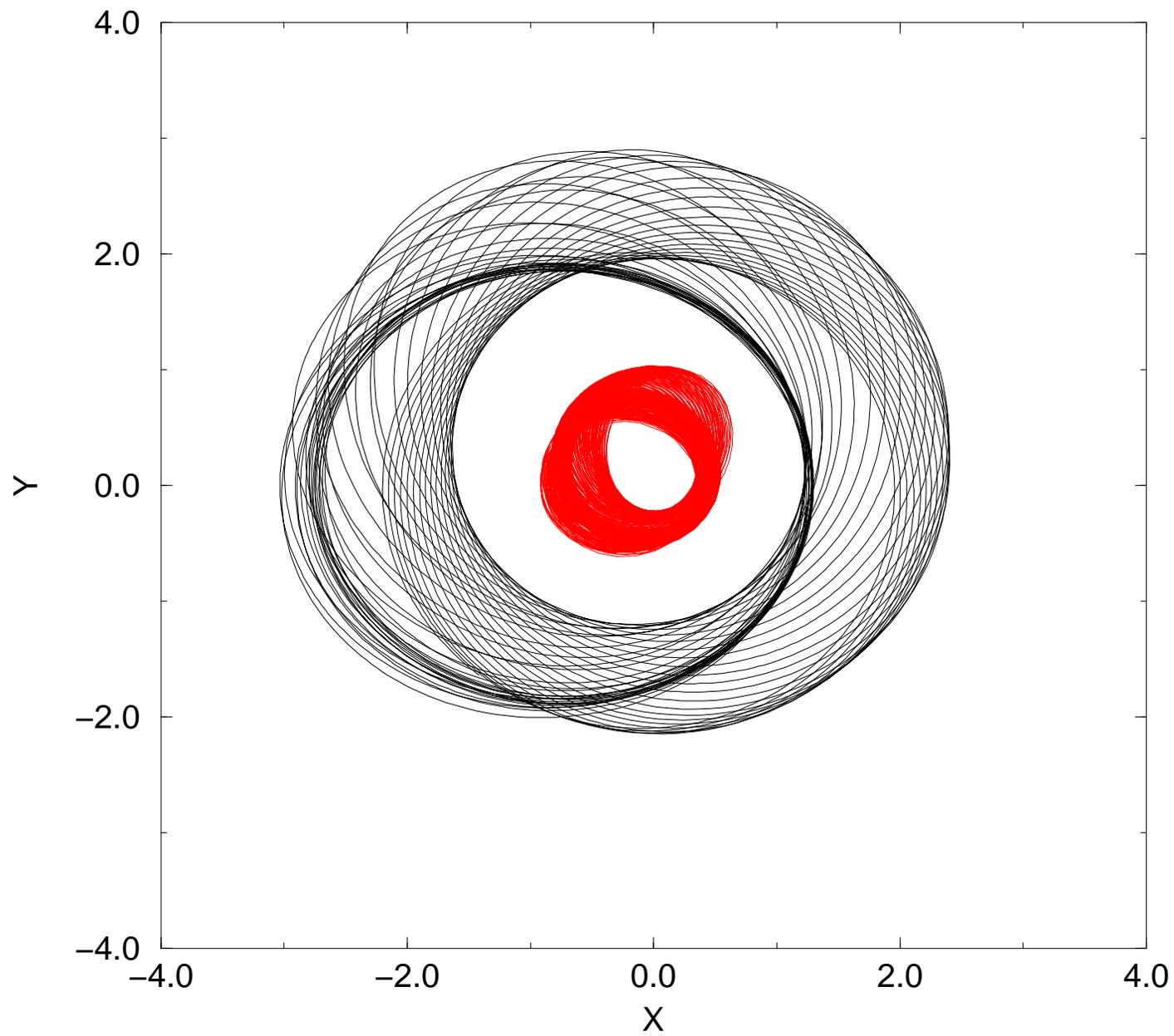
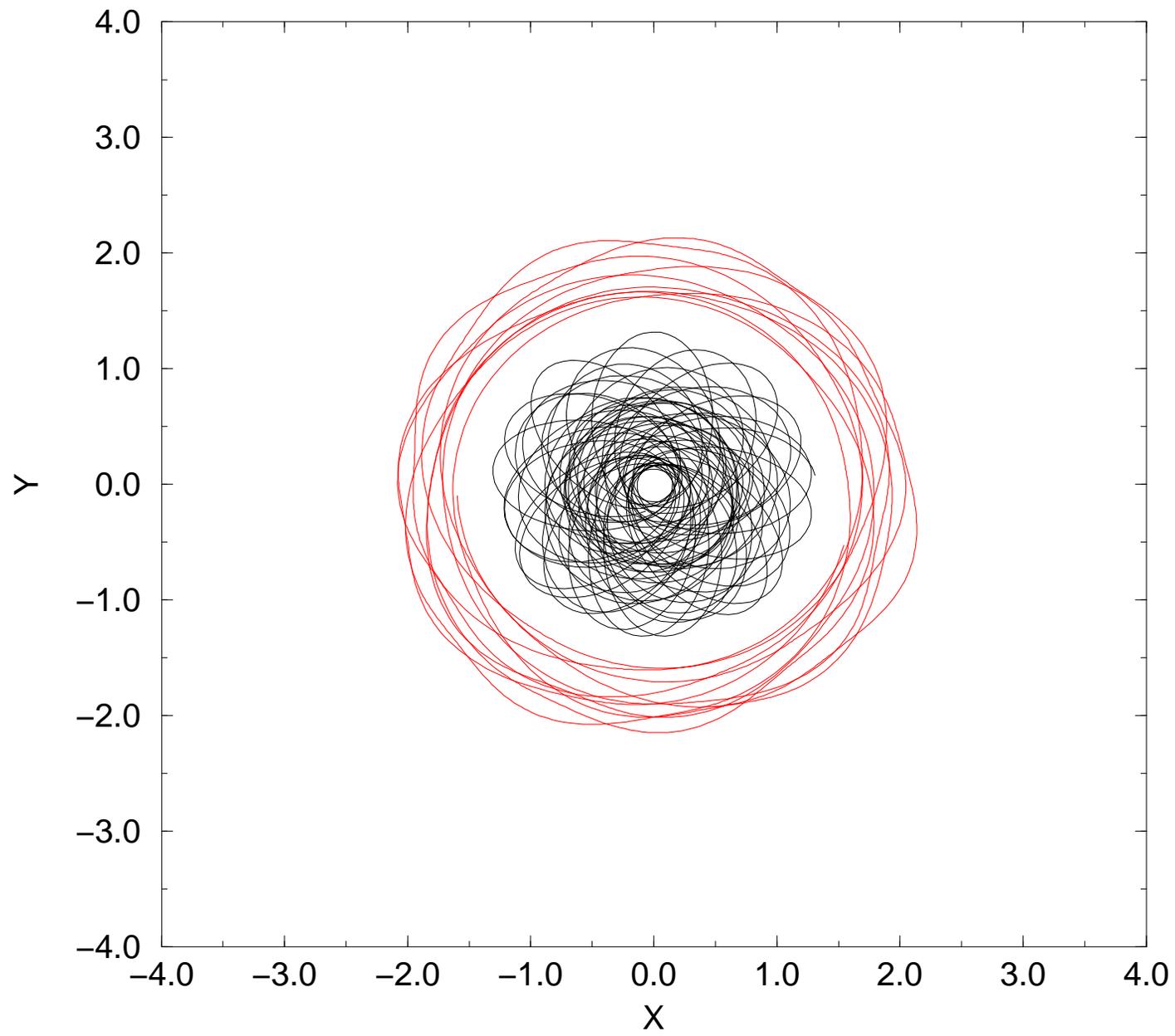
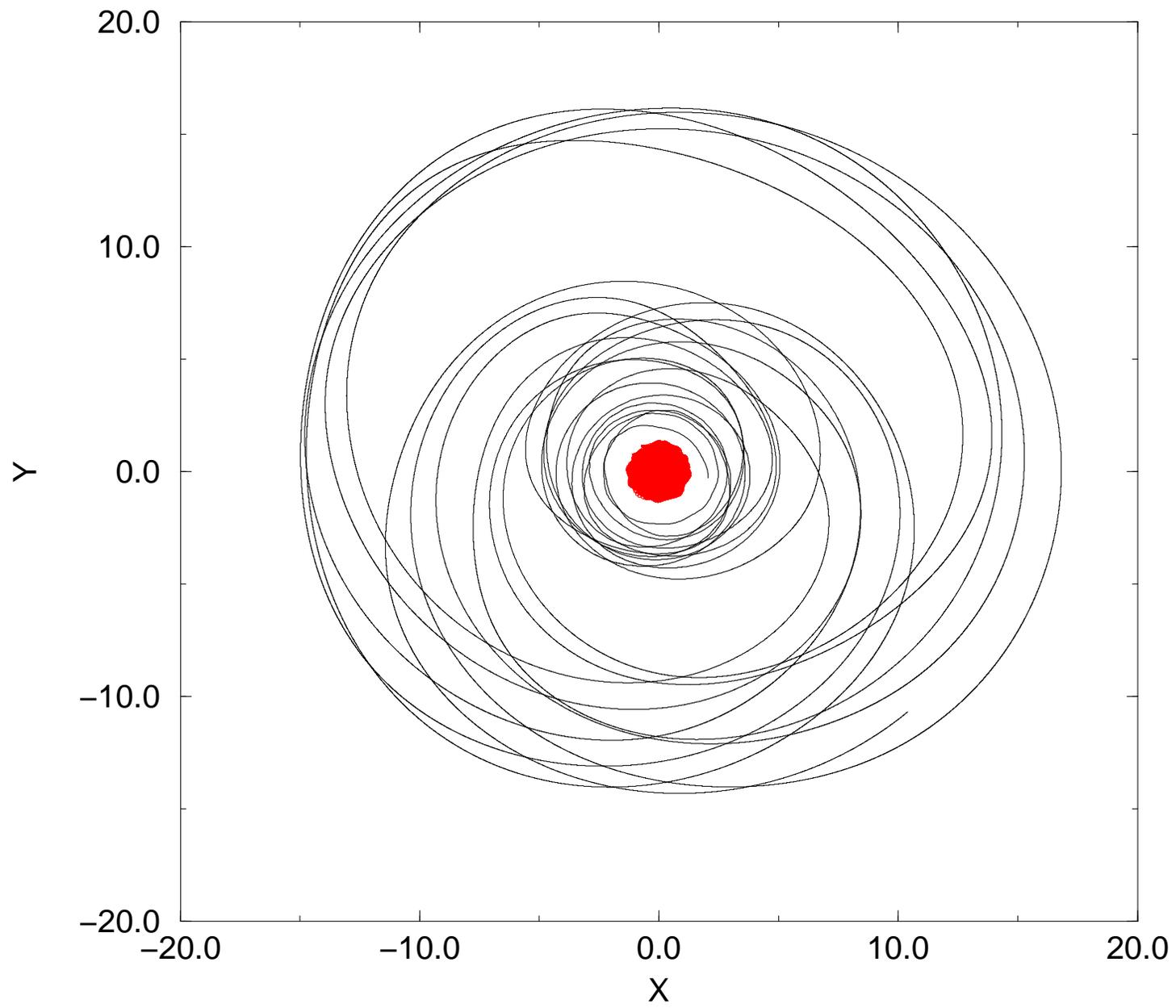
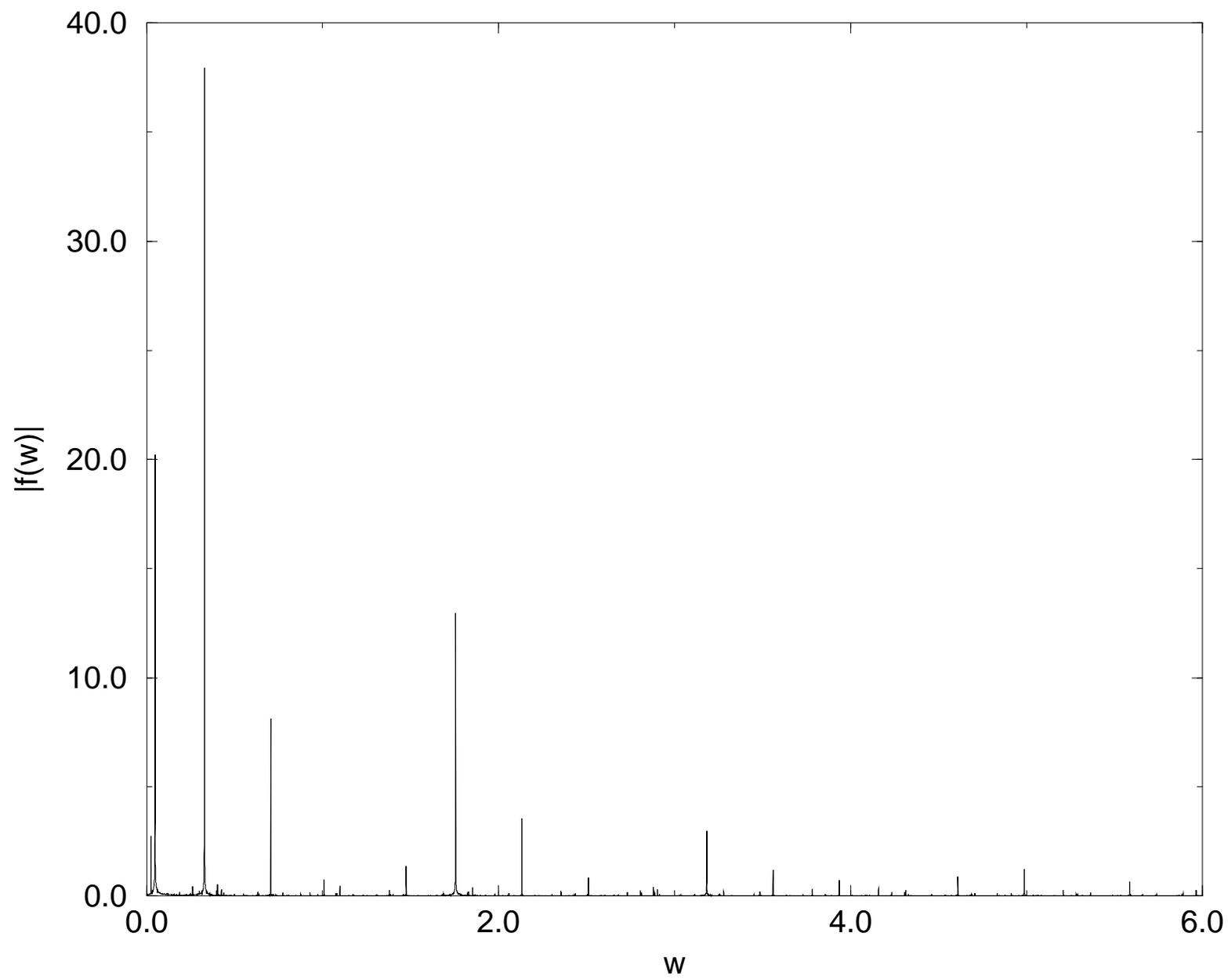
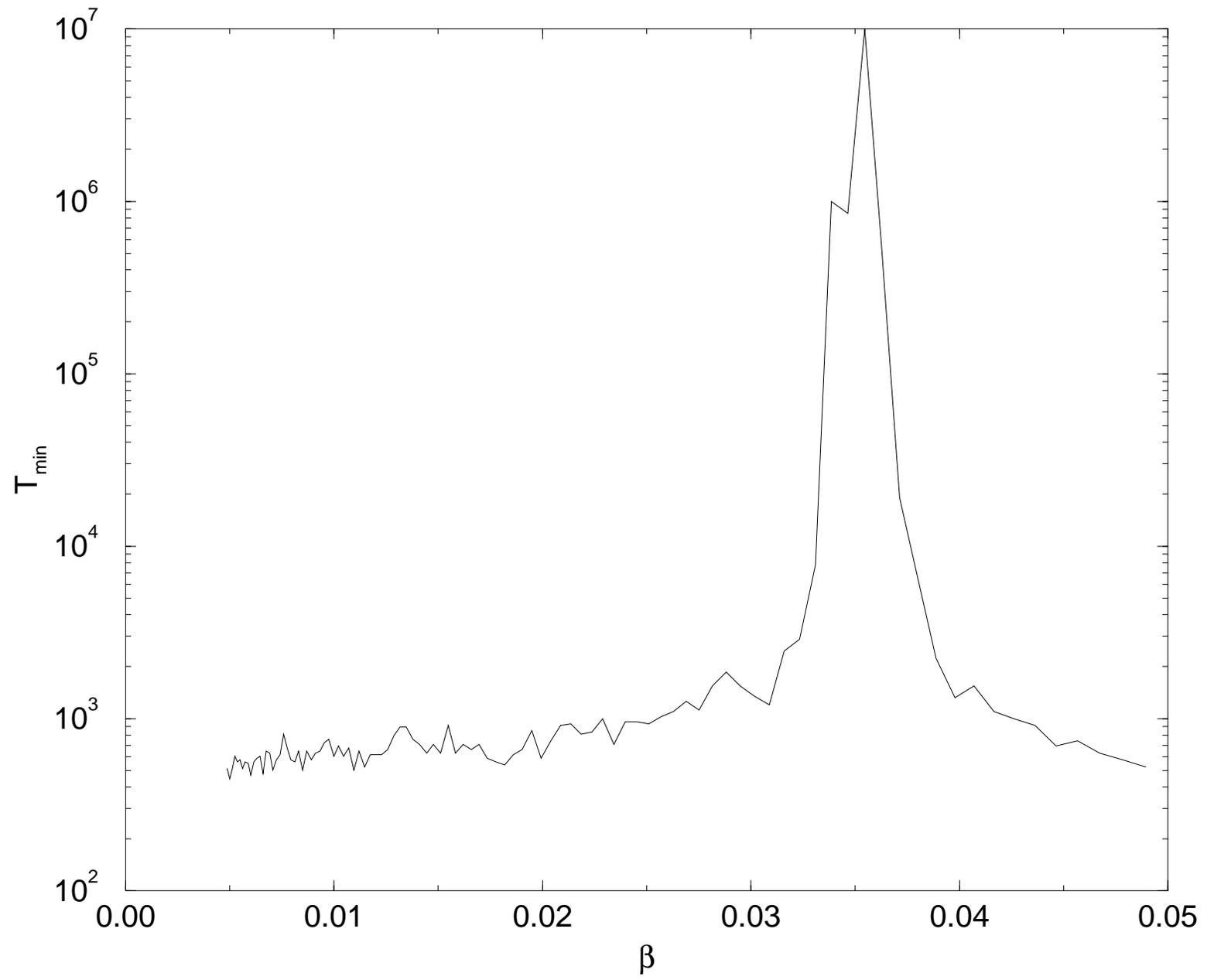
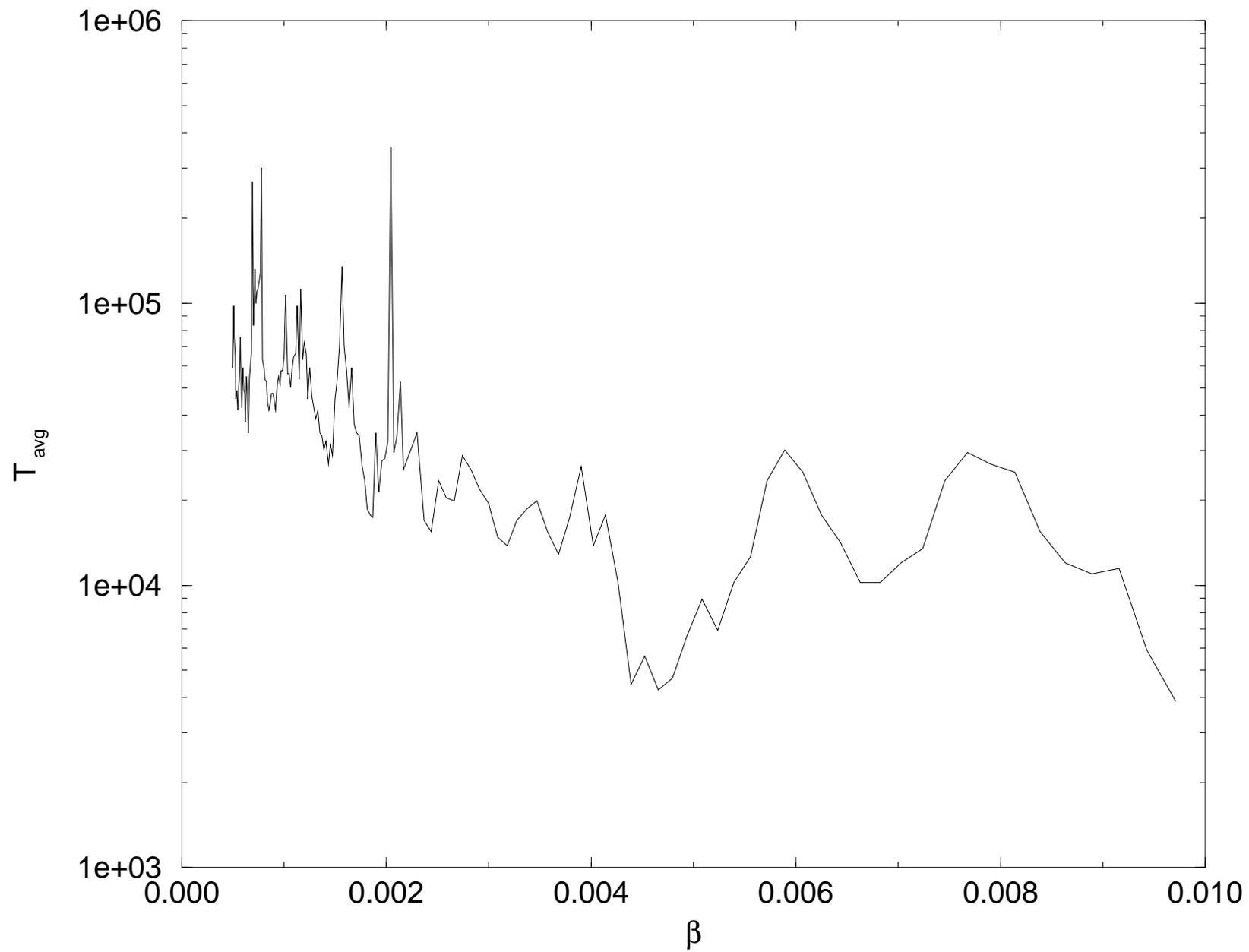
\end{document}